\documentclass[prl,twocolumn,aps,a4paper,superscriptaddress]{revtex4-1}
\usepackage{amsmath,mleftright,mathtools}
\usepackage{bm}
\usepackage{stix,microtype}
\usepackage{graphicx}
\usepackage{xspace}
\usepackage{units}
\usepackage[colorlinks,linkcolor=blue,citecolor=blue,urlcolor=blue]{hyperref}
\usepackage[dvipsnames]{xcolor}
\usepackage{cleveref}
\usepackage{array}
\usepackage{dcolumn}
\usepackage[normalem]{ulem}

\newcommand{\be}{\begin{equation}}
\newcommand{\ee}{\end{equation}}
\newcommand{\bea}{\begin{eqnarray}}
\newcommand{\eea}{\end{eqnarray}}

\def\a{\alpha}
\def\b{\beta}
\def\g{\gamma}
\def\G{\Gamma}
\def\d{\delta}
\def\D{\Delta}
\def\e{\epsilon}
\def\eps{\epsilon}

\def\h{\eta}
\def\th{\theta}

\def\m{\mu}
\def\n{\nu}

\def\p{\pi}

\def\r{\rho}
\def\s{\sigma}
\def\S{\Sigma}

\def\f{\phi}

\def\w{\omega}

\def\z{\zeta}

\def\bgr{\mbox{\boldmath $\rho$}}

\def\bcallh{\mbox{\boldmath $h$}}

\def\bcallv{\mbox{\boldmath $v$}}

\def\callG{\mbox{$\mathcal{G}$}}

\def\callI{\mbox{$\mathcal{I}$}}

\def\bcallG{\mbox{\boldmath $\mathcal{G}$}}

\def\bra{\langle}
\def\ket{\rangle}

\def\Tr{{\rm Tr}}
\def\Re{{\rm Re}}

\def\heff{h^e_{\text{eff}}}
\def\heffd{h^{e\dagger}_{\text{eff}}}
\def\im{{\rm i}}
\def\ud{{\rm d}}
\def\ex{{\rm e}}
\def\iu{{\rm i}}
\def\1op{\hat{\mathbbm{1}}}
\def\nn{\nonumber}

\begin{document}
\title{Time-linear quantum transport simulations with correlated nonequilibrium Green's functions}
\author{R. Tuovinen}
\affiliation{QTF Centre of Excellence, Department of Physics, P.O. Box 64, 00014 University of Helsinki, Finland \looseness=-1}
\affiliation{Department of Physics, Nanoscience Center, P.O. Box 35, 40014 University of Jyv{\"a}skyl{\"a}, Finland \looseness=-1}
\author{Y. Pavlyukh}
\affiliation{Institute of Theoretical Physics, Faculty of Fundamental Problems of Technology, Wroclaw University of Science and Technology, 50-370 Wroclaw, Poland}
\author{E. Perfetto}
\affiliation{Dipartimento di Fisica, Universit{\`a} di Roma Tor Vergata, Via della Ricerca Scientifica 1, 00133 Rome, Italy \looseness=-1}
\affiliation{INFN, Sezione di Roma Tor Vergata, Via della Ricerca Scientifica 1, 00133 Rome, Italy \looseness=-1}
\author{G. Stefanucci}
\affiliation{Dipartimento di Fisica, Universit{\`a} di Roma Tor Vergata, Via della Ricerca Scientifica 1, 00133 Rome, Italy \looseness=-1}
\affiliation{INFN, Sezione di Roma Tor Vergata, Via della Ricerca Scientifica 1, 00133 Rome, Italy \looseness=-1}
\begin{abstract}  
We present a time-linear scaling method to simulate open and correlated
quantum systems out of equilibrium. The method inherits from many-body perturbation 
theory the possibility to choose selectively
the most relevant scattering processes in the dynamics, 
thereby paving the way to the real-time characterization 
of correlated ultrafast phenomena in quantum transport. 
The open system dynamics is described in terms of an {\em embedding 
correlator} from which the time-dependent current can be calculated 
using the Meir-Wingreen formula. We show how to efficiently implement 
our approach through a simple grafting into recently proposed 
time-linear Green's function methods for closed systems. 
Electron-electron and electron-phonon interactions can be treated on 
equal footing while preserving all fundametal conservation laws.
\end{abstract}
\maketitle

{\em Introduction.--} Few systems in nature are in equilibrium. Behind the facade of, 
e.g., calm and stationary transport, the electrical and heat currents 
run violently. Such out-of-equilibrium dynamics bridges  
fields like quantum transport and optics~\cite{kuniyuki_many-body_2019, 
huebener_engineering_2021}, atomic and molecular physics~\cite{ruggenthaler_from_2018, 
mansson_real-time_2021}, spectroscopy in solids~\cite{buzzi_photomolecular_2020,dendzik_observation_2020, 
nicoletti_coherent_2022}, and cavity materials engineering~\cite{latini_the_ferroelectric_2021, 
bao_light-induced_2022, schlawin_cavity_2022}. Recent progress in state-of-the-art 
time-resolved pump-probe spectroscopy and transport measurements has 
pushed the temporal resolution down to the femtosecond 
time scale~\cite{mcIver_light-induced_2020, sung_long-range_2020, deSio_intermolecular_2021, abdo_variable_2021, 
niedermayr_few-femtosecond_2022}. Inherently, the associated phenomena are time-dependent; the 
complex many-body systems are far from equilibrium, with no guarantee 
of instantly relaxing to stationary states.

The theory of quantum transport 
began with the
pioneering works of Landauer and 
B\"uttiker~\cite{Landauer:57,buttiker_generalized_1985,buttiker_four-terminal_1986} and became a mature 
field after the works of Meir, Wingreen and 
Jauho~\cite{meir_landauer_1992,jauho_time-dependent_1994} who 
provided a general formula for the time-dependent current through 
correlated junctions in terms of nonequilibrium Green's functions 
(NEGF). NEGF is an \emph{ab initio} method suitable to deal with 
both bosonic and 
fermionic interacting 
particles, in and out of equilibrium~\cite{danielewicz_quantum_1984, 
svl-book, balzer_nonequilibrium_2013}. Nonetheless, the ability to 
harness the full power of the Meir-Wingreen formula~\cite{meir_landauer_1992} is hampered 
by the underlying two-time structure of the NEGF, a feature that 
makes real-time simulations computationally 
challenging.

In this Letter, we present a time-linear scaling NEGF theory for 
{\em open} and {\em correlated} quantum systems. 
The resulting method is strikingly simple, with 
ordinary differential equations (ODE) only. Correlation effects  
originating from different scattering mechanisms
are included through a proper selection of Feynman diagrams, 
and all fundamental conservation laws are preserved. The 
Meir-Wingreen formula is rewritten in terms of an {\em embedding 
correlator} which allows for evaluating the time-dependent current at a 
{\em time-linear} 
cost.
We use the method to study transport of 
electron-hole pairs, highlighting the pivotal role of 
correlations in capturing velocity renormalizations and decoherence 
mechanisms.

\begin{figure}[t]
\centering
\includegraphics[width=0.47\textwidth]{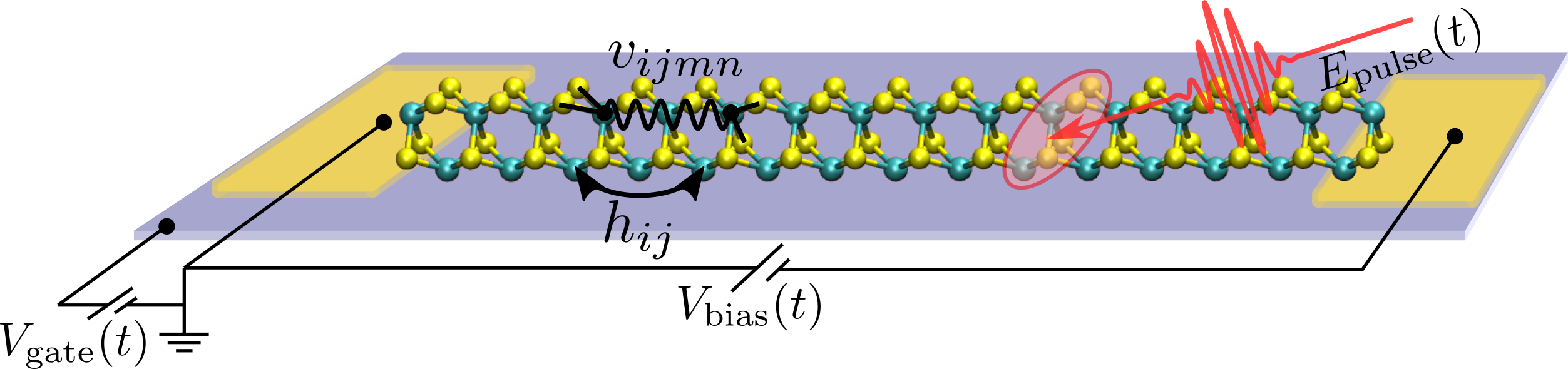}
\caption{Quantum transport set up. A 
nanowire (finite quantum system)
is contacted to left ($\a=L$) and right ($\a=R$) electrodes and lies over a substrate 
(or gate electrode) $\a=$gate. The nanowire is driven out of 
equilibrium by time dependent voltages $V_{\a}(t)$ [$V_{\rm bias}=V_{L}-V_{R}$] and laser pulses 
$E_{\rm pulse}(t)$.
}
\label{schematic-crop}
\end{figure}

{\em Kadanoff-Baym equations for open systems.--}
We consider a finite quantum system, being it a molecule or a 
nanostructure, with one-electron integrals $h_{ij}(t)$ and two-electron 
integrals $v_{ijmn}$ in some orthonormal one-particle basis of 
dimension $N_{\rm sys}$, 
see Fig.~\ref{schematic-crop}. The time-dependence of 
$h_{ij}(t)=h^{0}_{ij}+\bra i|V_{\rm 
gate}(t)+\hat{\mathbf{d}}\cdot{\mathbf{E}_{\rm pulse}}(t)|j\ket$ is 
due to a time-dependent gate voltage $V_{\mathrm{gate}}(t)$ and to 
a possible laser pulse $\mathbf{E}_{\rm pulse}(t)$ coupled to the 
electronic dipole operator $\hat{\mathbf{d}}$.
The system is said open if it is in contact with 
electronic reservoirs with which it can exchange particles and hence 
energy. This is the typical quantum transport set up, the finite 
system being the junction and the electronic reservoirs being the 
electrodes. Neglecting correlation effects in the electrodes and between the electrodes and the system
the Green's function 
$G_{ij}(z,z')$ with times $z,z'$ on 
the Keldysh contour $C$ satisfies the equation of motion (EOM) (in 
the $N_{\rm sys}\times N_{\rm sys}$ matrix 
form)~\cite{jauho_time-dependent_1994,jauho-book,mssvl.2008,mssvl.2009,svl-book}
\begin{align}
\left[\iu\frac{d}{dz}-h^{e}(z)\right]&G(z,z')
=\d(z,z')
\nn\\&+\!
\int_{C}d\bar{z}\;\big[\S_{\rm c}(z,\bar{z})+\
\S_{\rm em}(z,\bar{z})\big]G(\bar{z},z').
\label{gmmeq2}
\end{align}
In Eq.~(\ref{gmmeq2}) $h^{e}_{ij}(z)=h_{ij}(z)+V^{\rm HF}_{ij}(z)$
is the 
one-electron Hamiltonian 
properly renormalized by the Hartree-Fock (HF) potential $V^{\rm 
HF}_{ij}(z)=-i\sum_{mn}(v_{imnj}-v_{imjn})G_{nm}(z,z^{+})$,
$\S_{\rm c}$ is the correlation 
self-energy due to electron-electron interactions, 
and $\S_{\rm em}$ is the embedding self-energy accounting for all 
virtual processes where an electron from orbital $i$ leaves the 
system to occupy some energy level in one of the electrodes and 
thereafter moves back to the system in orbital $j$.

The Kadanoff-Baym equations (KBE) for the open system follow from 
Eq.~(\ref{gmmeq2}) by choosing the times $z$ and $z'$ on 
different branches of the Keldysh contour~\cite{svl-book}. 
In particular the EOM for 
the $N_{\rm sys}\times N_{\rm sys}$ 
electronic density matrix $\r(t)=-iG(z,z^{+})=-iG^{<}(t,t)$ can easily be 
derived by subtracting Eq.~(\ref{gmmeq2}) from its 
adjoint and then setting $z=t$ on the forward branch and $z'=t$ on 
the backward branch:
\begin{align}
  i\frac{d}{dt}\rho(t)=\Big(h^{e}(t)\rho(t) 
  -i I_{\rm c}(t)-i I_{\rm em}(t)\Big)-{\rm h.c.}.
  \label{eq:eomrho:e}
\end{align}
The 
collision integral $I_{\rm c}$ is the convolution between the 
correlation self-energy and the Green's function $G$ whereas the 
embedding integral $I_{\rm em}$, the main focus of this work, is 
the convolution between the embedding self-energy and $G$. For 
$I_{\rm em}=0$ the system is closed (no electrodes) and the KBE have 
been implemented in a number of works using different approximations to 
$\S_{\rm c}$; these include the second-Born~\cite{PhysRevLett.98.153004,BalzerHermanns2012}, 
the $GW$ and $T$-matrix~\cite{pva.2009,Stan2009,pva.2010,schuler_nessi_2020}, the Fan-Migdal~\cite{Niko2015,schuler_time-dependent_2016}, and approximations based 
on the nonequilibrium dynamical mean-field theory~\cite{Freericks2006, Aoki2014, Strand2015, Golez2017}.
KBE studies of open systems are less 
numerous~\cite{mssvl.2008,mssvl.2009,pva.2010}. 
In all cases the unfavorable 
$O(N_t^3)$ scaling with the number of time steps $N_t$
limits the KBE, and hence the possibility of studying 
ultrafast correlated dynamics,
to relatively small systems,
although promising progresses have been recently achieved~\cite{schuler_nessi_2020, Kaye2021, Meirinhos2022, Dong2022}.

{\em Generalized Kadanoff-Baym Ansatz.--}
For any given correlation self-energy
the direct solution of the EOM for the density matrix, see 
Eq.~(\ref{eq:eomrho:e}),  
is computationally less complex than solving the KBE
and opens the door to a wealth of nonequilibrium phenomena~\cite{PS-cheers}. 
To date the most efficient way to make the collision integral a 
functional of $\r$ is the Generalized 
Kadanoff-Baym Ansatz (GKBA)~\cite{lipavsky_generalized_1986}
\begin{align}
  G^{\lessgtr}(t,t')=-G^{R}(t,t')\rho^{\lessgtr}(t')+\rho^{\lessgtr}(t)G^{A}(t,t'),
  \label{eq:e:gkba}
\end{align}
with $G^{R/A}$ the $N_{\rm sys}\times N_{\rm sys}$ 
retarded/advanced quasi-particle propagators,
$\r^{>}=\r-1$ and $\r^{<}=\r$.
The GKBA respects the causal structure and it 
preserves all fundamental conservation laws for 
$\Phi$-derivable approximations~\cite{baym_self-consistent_1962}
to $\S_{\rm c}$~\cite{karlsson_fast_2021}. 
In {\em closed} systems the 
GKBA-EOM can be exactly reformulated in terms of a coupled set of ODEs~\cite{schlunzen_achieving_2020,joost_g1-g2_2020}
for several major approximations to 
$\S_{\rm c}$, the most notable being $GW$ and 
$T$-matrix~\cite{schlunzen_achieving_2020,joost_g1-g2_2020,perfetto_real_2022}, $GW$ and 
$T$-matrix plus 
exchange~\cite{pavlyukh_photoinduced_2021,pavlyukh_time-linear_2022}, 
Fan-Migdal~\cite{karlsson_fast_2021} and the doubly-screened 
$G\widetilde{W}$~\cite{pavlyukh_interacting_2022}. In essence, the 
idea is 
to introduce high-order correlators $\callG^{a=1,\ldots,n}(t)$ 
($\callG^{1}=\r$), write 
$I_{\rm c}[\{\callG^{a}\}]$ as a functional of them, and solve the 
coupled 
EOMs 
$i\frac{d}{dt}\callG^{a}=\callI^{a}[\{\callG^{a}\}]$ 
($\callI^{1}=\big(h^{e}\r-iI_{\rm 
c}\big)-{\rm h.c.}$). For all aforementioned methods the system 
of ODE can be closed using a relatively few number of  
correlators (the highest number being $n=7$ in $G\widetilde{W}$). 
Extending the ODE formulation to {\em open} systems would enable 
performing time-linear ($O(N_t)$ scaling) NEGF simulations of 
correlated junctions and hence studying, e.g., the formation of Kondo 
singlets~\cite{krivenko_dynamics_2019}, blocking dynamics of
polarons~\cite{albrecht_long_2013,wilner_nonequilibrium_2014}, 
bistability and hysteresis~\cite{galperin_hysteresis_2005,wilner_bistability_2013}, phonon dynamics and 
heating~\cite{galperin_molecular_2007,galperin_heat_2007,wilner_phonon_2014,rizzi_electron_2016}, 
nonconservative 
dynamics~\cite{todorov_current-induced_2001,dundas_current-driven_2009,bode_scattering_2011,feng_current-induce_2019},
molecular 
electroluminescence~\cite{kuniyuki_many-body_2019} as well as 
transport and optical response of junctions under periodic 
drivings~\cite{gabriel_optical_2020,xiao_kondo_2013}, see also 
Ref.~\cite{ridley_a_many-body_2022} for a recent review. 

Below we show that the set of ODEs for closed systems 
can be coupled to one more 
ODE for the {\em embedding correlator}  $\callG^{\rm em}$ 
to effectively open the system, thus 
providing a time-linear method to solve Eq.~(\ref{eq:eomrho:e}). 
Equation~(\ref{eq:eomrho:e}) was originally investigated using the integral 
(convolution) form of the collision 
and embedding integrals in Refs.~\cite{latini_charge_2014,Tuovinen2020,tuovinen_electronic_2021,tuovinen_electron_2021}. 
It was emphasized therein 
that the GKBA propagators $G^{R/A}$ chosen for closed-system simulations 
need to be modified. This change 
affects all other ODEs in an extremely elegant way while preserving 
the overall computational 
complexity.

{\em Time-linear method.--}
Let $\S_{\a}$ be the embedding 
self-energy of electrode $\a=1,\ldots,N_{\rm leads}$, 
hence $\S_{\rm em}=\sum_{\a}\S_{\a}$. 
In the so-called wide-band limit approximation 
(WBLA)~\cite{wblaobservation}, the 
retarded and lesser components read~\cite{svl-book,tuovinen_time-dependent_2014,suppmat}
\begin{subequations}
\begin{align}
	\S^{R}_{\a}(t,t')&=-\frac{i}{2}s^{2}_{\a}(t)\d(t,t')\,\G_{\a},
	\\
	\S^{<}_{\a}(t,t')&=is_{\a}(t)s_{\a}(t')e^{-i\f_{\a}(t,t')}\!\!
	\int\!\!\frac{d\w}{2\p}f(\w-\m)e^{-i\w(t-t')}\,\G_{\a},
\end{align}
\end{subequations}
where $s_{\a}(t)$ is the switch-on function for the contact 
between the system and electrode $\a$, $\G_{\a}$ is the
$N_{\rm sys}\times N_{\rm sys}$ quasi-particle line-width matrix due to electrode $\a$,
$\f_{\a}(t,t')\equiv \int_{t'}^{t}d\bar{t}\;V_{\a}(\bar{t})$ is the 
accumulated phase 
due to the time-dependent voltage $V_{\a}$~\cite{ridley_current_2015},
and $f(\w-\m)=1/(e^{\b(\w-\m)}+1)$ is the Fermi function at inverse 
temperature $\b$ and chemical potential $\m$. The matrix elements
$\G_{\a,ij}=2\p\sum_{k}T_{ik\a}\d(\m-\e_{k\a})T_{k\a j}$ can be 
calculated from the transition amplitudes $T_{k\a j}=T^{\ast}_{j k\a}$ 
from orbital $j$ to level $k$ in electrode $\a$ having the energy dispersion 
$\e_{k\a}$. The exact form of the 
embedding integral is then
\begin{align}
I_{\rm em}(t)=\sum_{\a}I_{\a}(t)=\int\!\! d\bar{t}\;
\S^{<}_{\rm em}(t,\bar{t})G^{A}(\bar{t},t)
+\frac{1}{2}\G(t)\r(t),
\label{embint}
\end{align}
with $\G(t)\equiv\sum_{\a}s^{2}_{\a}(t)\G_{\a}$. In 
Ref.~\cite{latini_charge_2014} it was shown that the mean-field approximation of 
Eq.~(\ref{eq:eomrho:e}), i.e., $I_{\rm c}=0$, is exactly reproduced 
in GKBA provided that 
\begin{align}
G^{R}(t,t')=-i\th(t-t')T 
e^{-i\int_{t'}^{t}d\bar{t}\big(h^{e}(\bar{t})-i\G(\bar{t})/2\big)},
\label{gret}
\end{align}
and $G^{A}(t',t)=[G^{R}(t,t')]^{\dag}$. 
Equation~(\ref{gret}) reduces to the propagator of closed 
systems for $\G=0$. In open systems, however, setting  $\G=0$ is 
utterly inadequate as no steady-state would ever be attained.
Beyond the mean-field approximation we close 
Eq.~(\ref{eq:eomrho:e}) using the GKBA with propagators 
as in Eq.~(\ref{gret})~\cite{propmisc}.

To construct the time-linear method we use an efficient 
pole expansion (PE) scheme for the Fermi function~\cite{hu_communication_2010} 
$f(\w)=\frac{1}{2}- \sum_{l} \eta_l \left(\frac{1}{\b 
\w+i\zeta_l}+\frac{1}{\b \w -i\zeta_l}\right)$, $\Re[\z_{l}]>0$, to rewrite the lesser 
self-energy for $t>t'$ as $\Sigma_\a^<(t,t')=\frac{i}{2} s_{\a}^2(t)\delta(t-t')\Gamma_\a
-s_{\a}(t)\sum_{l}\frac{\eta_l}{\beta}F_{l\a}(t,t')\G_{\a}$ with
\begin{align}
F_{l\a}(t,t') =  s_{\a}(t')e^{-i\f_\a(t,t')}
e^{-i(\mu-i\frac{\zeta_l}{\beta})(t-t')}.
\label{defFla}
\end{align}
Inserting the result into Eq.~(\ref{embint}) the EOM Eq.~(\ref{eq:eomrho:e}) for the density 
matrix becomes
\begin{align}
  i\frac{d}{dt}\rho&=\Big(h^{e}_{\rm eff}\rho+\frac{i}{4}\G+i\sum_{l\a}
  s_{\a}\frac{\eta_l}{\beta}\G_{\a}\callG^{\rm em}_{l\a}-i I_{\rm c}\Big)  
  -{\rm h.c.},
  \label{gkbaoseom}
\end{align}
where $h^{e}_{\rm eff}(t)\equiv h^{e}(t)-i\G(t)/2$ is the effective 
(non-self-adjoint) mean-field Hamiltonian and $\callG^{\rm em}_{l\a}(t)\equiv \int\!\! 
d\bar{t}\;F_{l\a}(t,\bar{t})G^{A}(\bar{t},t)$ is the $N_{\rm sys}\times N_{\rm sys}$
embedding correlator. Taking into account the explicit expressions in
Eqs.~(\ref{gret}) and (\ref{defFla}) we find
\begin{align}
\!\! i\frac{d}{d t} \callG^{\rm em}_{l\a}(t)  = - s_{\a}(t) - 
 \mathcal{G}^{\rm em}_{l\a}(t)\Big(h^{e\dag}_{\rm 
eff}(t)-V_\a(t)-\mu+i\frac{\zeta_l}{\beta}\Big).
\label{eomgla}
\end{align}
Equations (\ref{gkbaoseom}) and (\ref{eomgla}), together with the 
ODEs for $I_{\rm c}$, form a coupled system of ODEs for correlated 
real-time simulations of open systems. 
This time-linear method
becomes similar to the one of 
Refs.~\cite{croy_propagation_2009,xiao_time-dependent_2010,yu_first-principles_2013,kwok_time-dependent_2013,wang_time-dependent_2013,kwok_stm_2019} 
for $I_{\rm c}=0$. The scaling with the system size of 
Eq.~(\ref{eomgla}) grows like $N_{\rm sys}^{3}\times N_{p}\times N_{\rm leads}$ 
where $N_{p}$ is the number 
of poles for the expansion of $f(\w)$~\cite{suppmat}.

An alternative time-linear method can be 
constructed from the spectral decomposition (SD) of the embedding self-energy
$\S_{\a,ij}(z,z')=s_{\a}(z)s_{\a}(z')\sum_{k}T_{i k\a}g_{k\a}(z,z')T_{k\a j}$, where 
$g_{k\a}$ is the Green's 
function of the isolated electrode. In this case, one would rewrite 
the embedding integral as $I_{{\rm 
em},ij}=\sum_{k\a}T_{i k\a}\widetilde{\callG}^{\rm em}_{k\a j}$ and derive an 
ODE for the scalar quantities
$\widetilde{\callG}^{\rm em}_{k\a j}=\sum_{m}\int 
d\bar{t}[g^{R}_{ka}(t,\bar{t})T_{ka m}G^{<}_{mj}(\bar{t},t)+
g^{<}_{ka}(t,\bar{t})T_{k\a m}G^{A}_{mj}(\bar{t},t)]$ 
using the GKBA for the lesser Green's function. 
The scaling with the system size of the 
scalar ODE for all $\widetilde{\callG}^{\rm em}_{k\a j}$
grows like $N_{\rm sys}^{2}\times N_{k}\times N_{\rm leads}$~\cite{suppmat}, where 
$N_{k}$ is the number of  $k$-points needed 
for the discretization.
The SD scheme is ill-advised for the following reasons. 
If the electrodes are not wide band then the 
calculation of the mean-field propagator 
scales cubically in time; any other approximation to $G^{R}$, 
including Eq.~(\ref{gret}), would be inconsistent and could even lead to 
unphysical time evolutions, e.g., no steady states for 
constant voltages. If the electrodes are 
wide band then $N_{k}$ could be 
orders of magnitude larger than $N_{\rm sys}\times N_{p}$
to achieve convergence, hence $N_{\rm sys}^{2}\times N_{k}\gg N_{\rm 
sys}^{3}\times N_{p}$. This statement is proven numerically 
below; see also Supplemental Material~\cite{suppmat}.

The quasi-particle broadening $\G$ in the propagators, see 
Eq.~(\ref{gret}), is only responsible for a minor change in the ODEs 
for the high-order correlators of closed systems. We focus here on the 
$T$-matrix approximation in the particle-hole channel ($T^{ph}$) as 
$T^{ph}$-simulations of open systems are reported below; similar 
arguments apply to all other approximations in Ref.~\cite{pavlyukh_interacting_2022}. 
The collision integral is $I_{{\rm 
c},ij}=-i\sum_{lmn}v_{inml}\callG^{\rm c}_{lmjn}$ where $\callG^{\rm c}_{lmjn}=-
\bra\hat{d}^{\dag}_{j}\hat{d}^{\dag}_{n}\hat{d}_{l}\hat{d}_{m}\ket_{\rm c}$ is 
the correlated part of the equal-time two-particle Green's 
function~\cite{svl-book}.
Following Refs.~\cite{pavlyukh_photoinduced_2021,pavlyukh_time-linear_2022}, we construct the 
matrices in the two-electron space $\bcallG^{\rm c}_{\begin{subarray}{c}ij\\ 
nm\end{subarray}}(t)\equiv \callG^{\rm c}_{imjn}(t)$,  
$\bcallv_{\begin{subarray}{c}ij\\ 
nm\end{subarray}}\equiv v_{imnj}$ and 
$\bgr^{\lessgtr}_{\begin{subarray}{c}ij\\ nm\end{subarray}}\equiv 
\rho^{\lessgtr}_{ij}\rho^{\gtrless}_{mn}$ 
(boldface letters are used 
to distinguish them from matrices in one-electron space). 
The only difference in the derivation of the 
EOM for $\bcallG$ of closed systems~\cite{pavlyukh_photoinduced_2021,pavlyukh_time-linear_2022}
comes from the fact that the EOM for the propagator 
contains $h^{e}_{\rm eff}$ instead of $h^{e}$. The final result is 
therefore
\begin{align}
  i \frac{d}{dt}\bcallG^{\rm c}=\Big(\bgr^{<}\bcallv\bgr^{>}
  +\big(\bcallh^{e}_{\rm eff} 
   +\bgr^\Delta\bcallv\big)\bcallG^{\rm c}\Big)-{\rm h.c.},
   \label{eq:G2:X:init}
\end{align}
where $\bgr^\Delta=\bgr^{>}-\bgr^{<}$ and 
${\bcallh^{e}_{\mathrm{eff},}}_{\begin{subarray}{c}ij\\nm\end{subarray}}
\equiv h^{e}_{{\rm eff},ij}\delta_{mn}-\delta_{ij}h^{e\dag}_{{\rm eff},nm}$. 
The 
solution of the 
coupled  ODEs for 
$\r$, $\callG^{\rm em}$ and $\callG^{\rm c}$ yields the time-dependent 
evolution of open systems in the $T^{ph}$ approximation. Similarly, 
one can show that all the 2$^{12}$ NEGF methods of 
Ref.~\cite{pavlyukh_interacting_2022} are only affected by the 
replacement $h^{e}\to h^{e}_{\rm eff}$. The addition of $I_{\rm 
em}$ in the EOM for $\r$ along with the propagation of the embedding correlator according to Eq.~(\ref{eomgla}), allows for studying
open systems for a large number of NEGF 
methods. These include methods to deal with the electron-phonon 
interaction as well~\cite{karlsson_fast_2021}. 
Noteworthily, all NEGF methods in 
Ref.~\cite{pavlyukh_interacting_2022}
guarantee the satisfaction of 
fundamental conservation laws like the continuity equation and the 
energy balance.

{\em Charge current.--}
Charge distributions, local currents, local moments, etc., can be 
extracted from the one-particle density matrix $\r$. Information on 
the electron-hole pair correlation function is carried by the 
$T^{ph}$ correlator $\callG^{\rm c}$. The embedding 
correlator $\callG^{\rm em}$ is instead crucial for calculating the time-dependent current 
$J_{\a}(t)$ at the interface between the system and electrode $\a$. 
This current is given by the Meir-Wingreen formula~\cite{meir_landauer_1992} 
and it can be written as the contribution of the $\a$ electrode to 
the embedding integral~\cite{jauho_time-dependent_1994,jauho-book,svl-book}, see 
Eq.~(\ref{embint}), $J_{\a}(t)=2\Re\Tr[I_{\a}(t)]$. Expressing 
the embedding self-energy in terms of $\callG^{\rm em}$ we find 
\begin{align}
\!\!\! J_{\a}(t)=2s_{\a}(t)\Re\Tr\Big[\G_{\a}\Big(\!
s_{\a}(t)\frac{2\r(t)-1}{4}-\!\sum_{l}\frac{\h_{l}}{\b}\callG^{\rm em}_{l\a}(t)\!\Big)
\Big].
\end{align}
Satisfaction of the continuity equation 
implies $\mathrm{CE}\equiv\frac{d}{dt}\Tr[\r]+\sum_{\a}J_{\a}=0$.

{Spectral decomposition vs pole expansion.--}
We first study a two-level molecule 
coupled to two one-dimensional tight-binding electrodes. We set $h_{11}=h_{22}=0$, 
$h_{12}=h_{21}=-T/2$ and measure all energies in units of $T>0$. We consider an
interaction 
$v_{ijmn}=v_{ij}\delta_{in}\delta_{jm}$ with $v_{11}=v_{22}=1$ and 
$v_{12}=v_{21}=0.5$.
The chemical potential is fixed at the middle of the HF gap of the 
uncontacted system, in our case $\mu=1$, and the inverse temperature 
is $\beta=100$. The electrodes are parameterized by  
on-site and hopping energies $a_\a=\m$ (half-filled electrodes), 
$b_\a=-8$, respectively, the energy 
dispersion thus taking the form $\e_{k\a}=a_\a - 2|b_\a|\cos[\pi 
k/(N_k+1)]$, with $N_{k}$ the number of discretized $k$ points. 
The left and right electrodes are coupled to the first and second 
molecular levels, respectively, with transition amplitude 
$T_\a=-0.2$, $\a=L,R$. As $T_{\a}\ll b_{\a}$  
the WBLA is accurate and 
one finds $\G_{L,ij}=\d_{i1}\d_{j1}\g_{L}$ 
and $\G_{R,ij}=\d_{i2}\d_{j2}\g_{R}$ with 
$\g_{\a}=2T^{2}_{\a}/|b_{\a}|=0.01$.        

\begin{figure}[t]
\centering
\includegraphics[width=0.47\textwidth]{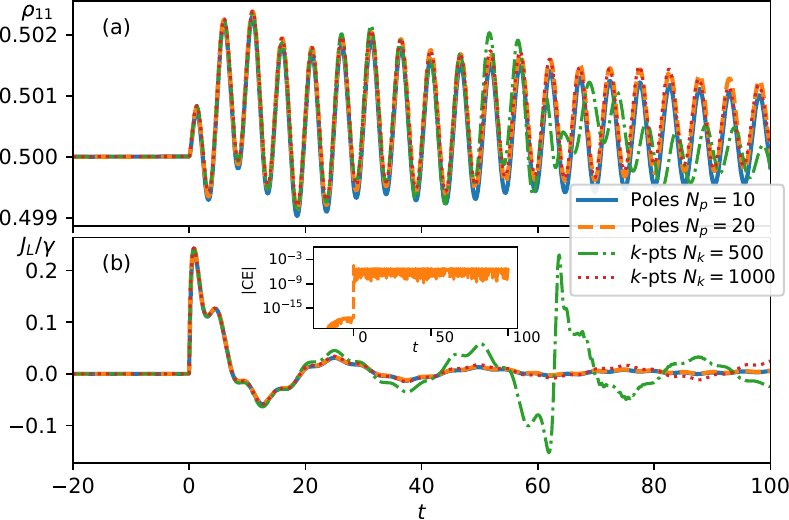}
\caption{Dynamics of the two-level molecule contacted to a left and 
right electrodes. Occupation of the first level (a) 
and current at the left interface (b) using the 
SD ($k$-points $N_{k}$) and PE (poles $N_{p}$) schemes. The inset in panel (b) 
shows the continuity equation $\mathrm{CE}=\frac{d}{dt}\mathrm{Tr}[\rho]+\sum_\a J_\a$.
Energies in units of $T$ and time in units of $1/T$.
}
\label{fig:dimer}
\end{figure}

In Fig.~\ref{fig:dimer} we present time-dependent HF results for 
the occupation of the first level [panel (a)] and the current at the left 
interface [panel (b)]. We adiabatically switch on the contacts 
between the molecule and the electrodes for $t<0$ with a 
``sine-squared'' switch-on function~\cite{KvLPS.2018}, and then drive the system away from 
equilibrium with a 
constant bias $V_L=T/2$, $V_R=0$ for $t\geq 0$ 
(hence $V_{\mathrm{gate}}=0$, $E_{\mathrm{pulse}}=0$, 
$V_{\mathrm{bias}}=V_L-V_R$).
The time-linear PE and SD schemes perform 
similarly at convergence, as they should. However, within 
the time-frame of the simulation, $N_{k}=1000$ $k$-points are needed to 
converge the SD scheme, against only $N_{p}=20$ 
poles needed to 
converge the PE scheme. Furthermore, the convergence with 
$N_{p}$ is independent of the maximum simulation time whereas $N_{k}$ must 
grow linearly with it for otherwise finite size effects, as those 
 visible for $N_{k}=500$ at time $t\simeq 50$, 
take place. 
Steady values are attained on a time scale of a few 
$1/\g_{a}$ time units~\cite{suppmat}. 
The inset in Fig.~\ref{fig:dimer}(b) shows 
that the continuity equation is satisfied with high accuracy.

{\em Correlated electron-hole transport.--} Transport of 
correlated electron-holes ($eh$) is a fundamental process in photovoltaic 
junctions~\cite{ponseca_ultrafast_2017,pastor_situ_2019}. 
We study the relaxation of a suddenly created $eh$ in a two-band direct gap one-dimensional semiconductor coupled to WBLA electrodes. 
The Hamiltonian of the system reads 
$\hat{H}= 
\sum_{ij\nu}h_{ij\nu}\hat{d}_{i\nu}^\dagger\hat{d}_{j\nu} + 
U\sum_{i}\hat{n}_{iv}\hat{n}_{ic}$,
where $\hat{d}_{i\nu}$ destroys an electron in the $i$-th valence 
($\n=v$) or conduction ($\n=c$) orbital, and 
$\hat{n}_{i\n}\equiv \hat{d}_{i\n}^\dagger\hat{d}_{i\n}$ is the 
orbital occupation. The one-electron integrals 
are $h_{ii v} = -\e_0<0$, $h_{ii c}=\e_{0}-U$
on site and $h_{ijv}=-h_{ijc}=T>0$ for nearest 
neighbors~\cite{perfetto_pump-driven_2019}. In equilibrium the HF gap 
is $\D=2(\e_{0}-2T)$. The 
left and right electrodes are coupled 
to the left- and right-most orbitals, respectively, with tunneling 
strength $\gamma_\a$ independent of $\a$.
Henceforth all energies are measured in 
units of $\D/2$; we set $\e_{0}=4.5$, $T=1.75$, $\g_{\a}=0.1$ and 
work at inverse temperature $\beta=100$.
The equilibrium chemical potential is set 
in the middle of the HF gap of the uncontacted system.

\begin{figure*}[tbp]
\centering
\includegraphics[width=0.95\textwidth]{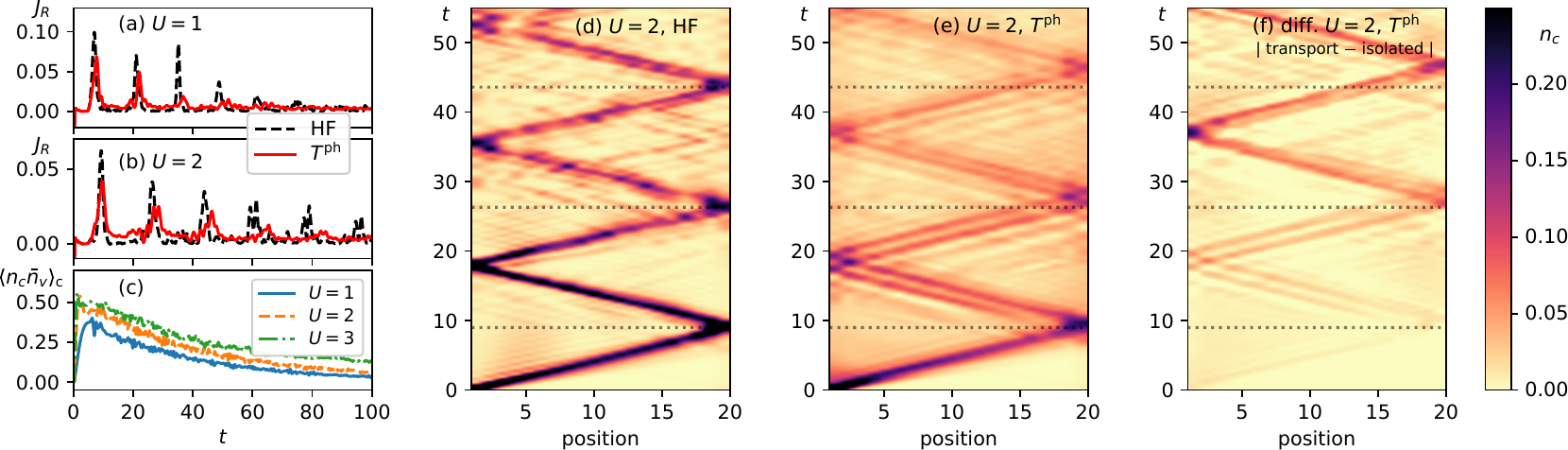}
\caption{Dynamics of an electron-hole pair in a one-dimensional 
semiconductor junction with 
$N=20$ cells. (a-b) Time-dependent current at the right-interface 
in HF (dashed) and $T^{ph}$ (solid) for $U=1$ (a) and $U=2$ (b). (c) 
Correlated part of the total number of $eh$ pairs for different 
interaction strengths. (d-e) Conduction 
occupations (color map) versus time (vertical axis) and cell position (horizontal axis) in HF 
(d) and $T^{ph}$~(e). (f) Difference of panel (e) to a situation without electrodes. The dashed lines are guides to the eye.}
\label{evolution}
\end{figure*}

At time $t=0$ we suddenly couple the system to the electrodes and 
create an $eh$ excitation at the left-most orbitals, hence
$\rho_{iv,iv}(0)=(1-\d_{i1})$ and $\rho_{ic,ic}(0)=\d_{i1}$. In 
Fig.~\ref{evolution}(a) and (b), we show the current at the right 
interface in two different many-body methods, i.e., HF and $T^{ph}$, and 
for two values $U=1,2$ of the $eh$ attraction. The results 
indicate that: (i) the velocity of the $eh$ wavepacket is faster in HF than 
$T^{ph}$  (each spike corresponds to an $eh$ bouncing at the 
right interface); (ii) the HF dynamics is coherent, the wavepacket 
travelling almost undisturbed, whereas in $T^{ph}$ correlations are 
responsible for a fast decoherence and wavepacket spreading. 
The slower velocity in $T^{ph}$ is rationalized in 
Fig.~\ref{evolution}(c) where we plot the correlated part of the 
total number of $eh$ pairs:   
$\sum_{i}\bra(1-\hat{n}_{iv})\hat{n}_{ic}\ket_{\rm c}=-\sum_{i}\callG^{\rm 
c}_{icivivic}$. The build-up of correlations is almost instantaneous. 
The initially uncorrelated $eh$ pair binds and becomes heavier, 
thus reducing the propagation speed. The observed decay at longer 
times is due to electron and hole tunneling into the electrodes; at the steady 
state about $10^{-2}$ conduction electrons and valence holes 
remain in the system 
(not shown). 
Both decoherence and velocity reduction are well visible in 
Fig.~\ref{evolution}(d) and (e) where we display the color plot of the 
conduction occupations $n_{ic}(t)=\rho_{ic,ic}(t)$ in HF and $T^{ph}$ 
for $U=2$. In $T^{ph}$ the $eh$ wavepacket loses coherence and 
spreads after bouncing back and forth a few times. In Fig.~\ref{evolution}(f) we analyze the effect of the electrodes 
by showing the difference between the open and closed system dynamics.
In the open system the amplitude of the 
wavepacket and the localization of the charge 
decreases faster than in the isolated system.

In conclusion, we put forward a time-linear approach to study the 
correlated dynamics of open systems with a large number of NEGF 
methods. Our work empowers the Meir-Wingreen formula 
allowing its use in 
contexts and/or for levels of approximation which were  
previously unattainable in practice. The ODE formulation lends 
itself to parallel computation, adaptive time-stepping implementations and 
restart protocols,  thus opening new avenues for the {\em ab initio}
description of time-dependent quantum transport phenomena.

\acknowledgments
R.T. wishes to thank the Academy of Finland for funding under Project No. 345007.
%Y.P. acknowledges funding from NCN Grant POLONEZ BIS 1, ``Nonequilibrium 
%electrons coupled with phonons and collective orders'', 
%2021/43/P/ST3/03293.   
Y.P. acknowledges funding from Project No. 2021/43/P/ST3/03293 co-funded by the National Science Centre and the European Union Framework Programme for Research and Innovation, Horizon 2020, under the Marie Sklodowska-Curie Grant Agreement No. 945339.
G.S. and E.P. acknowledge funding from MIUR PRIN Grant
No. 20173B72NB, from the INFN17-Nemesys project.
G.S. and E.P.  acknowledge
Tor Vergata University for financial
support through projects ULEXIEX and TESLA.
We also acknowledge CSC -- IT Center for Science, 
Finland, for computational resources.


%%%%%%%%%% Supplemental material begins here %%%%%%%%%%
\newpage
\begin{widetext}
\begin{center}
\textbf{\large Supplemental Material for ``Time-linear quantum transport simulations with correlated nonequilibrium Green's functions''}
\end{center}
\setcounter{equation}{0}
\setcounter{figure}{0}
\setcounter{table}{0}
\setcounter{page}{1}
In this Supplemental Material, we consider only the embedding self-energy between the system and the electrodes. Other collision terms, such as electron-electron and electron-phonon interactions, can be taken care of by a separate calculation~\cite{karlsson_fast_2021supp, pavlyukh_time-linear_2022supp}.
\end{widetext}

\section{Quantum transport setup}

We consider the following Hamiltonian for the quantum-correlated system coupled to electrodes
\begin{align}\label{eq:ham}
\hat{H} & = \sum_{k\a,\s} \eps_{k\a}\hat{d}_{k\a,\s}^\dagger \hat{d}_{k\a,\s} + \sum_{ij,\s}h_{ij}\hat{d}_{i,\s}^\dagger\hat{d}_{j,\s} \nonumber \\
& + \sum_{ik\a,\s}\left(T_{ik\a}\hat{d}_{i,\s}^\dagger\hat{d}_{k\a,\s}+\text{h.c.}\right) \nonumber \\
& +\frac{1}{2}\sum_{\substack{ijmn\\ \s\s'}}v_{ijmn}\hat{d}_{i,\s}^\dagger\hat{d}_{j,\s'}^\dagger\hat{d}_{m,\s'}\hat{d}_{n,\s} ,
\end{align}
where $\hat{d}^{(\dagger)}$ are the electronic annihilation (creation) operators, $\eps_{k\a}$ is the energy dispersion of the $\a$-th electrode, $h_{ij}$ are the one-particle matrix elements of the system, $T_{ik\a}$ are the tunneling matrix elements between the system and the electrodes, and $v_{ijmn}$ are the Coulomb integrals of the system. An out-of-equilibrium condition, making charge carriers to flow through the system, is introduced by assigning time-dependence on the horizontal branch of the Keldysh contour, $\eps_{k\a} \to \eps_{k\a}+V_\a(t) \equiv \bar{\eps}_{k\a}(t)$, $h_{ij} \to h_{ij}(t)$, and $T_{ik\a} \to T_{ik\a}s_\a(t)$, where $V_\a(t)$ is the (time-dependent) bias-voltage profile, and $s_\a(t)$ is a switch-on function for the system-electrode coupling. While the two-body interaction is itself instantaneous, we could also ramp the strength of the Coulomb integrals with a switch-on function.

\section{Retarded self-energy and the effective Hamiltonian}

The retarded self-energy appears in the equation of motion for the retarded Green's function:
\be\label{eq:gr}
\left[\im\frac{\ud}{\ud t} - h^e(t)\right]G^{\text{R}}(t,t') = \delta(t,t') + \int \ud \bar{t} \Sigma^{\text{R}}(t,\bar{t})G^{\text{R}}(\bar{t},t'),
\ee
where $h^e(t)$ is the (time-dependent) one-electron Hamiltonian, including the Hartree-Fock potential. The self-energy is constructed from the tunneling matrices and the electrode Green's function as $\Sigma^{\text{R}}(t,t') = \sum_\a \Sigma^{\text{R}}_\a(t,t')$ with~\cite{tuovinen_time-dependent_2014supp}
\be\label{eq:sigmaretk}
\Sigma_{\a,ij}^{\text{R}}(t,t') = \sum_k T_{ik\a}s_\a(t)g_{k\a}^{\text{R}}(t,t')T_{k\a j}s_\a(t'),
\ee
where we assumed the tunneling strength between the system and 
electrode $\a$ depends on time only through an overall switch-on 
function $s_\a(t)$. The free-electron, retarded Green's function of the $\a$-th electrode is
\be\label{eq:gretk}
g_{k\a}^{\text{R}}(t,t') = -\im\theta(t-t')\ex^{-\im\phi_\a(t,t')}\ex^{-\im\eps_{k\a}(t-t')}
\ee
with $\eps_{k\a}$ and $\phi_\a(t,t') \equiv \int_{t'}^t \ud \bar{t}V_\a(\bar{t})$ 
the energy dispersion and the bias-voltage phase factor, respectively. Inserting this into the expression for the self-energy, we can transform the $k$-summation into a frequency integral as
\begin{align}\label{eq:sigmaret}
\Sigma_{\a,ij}^{\text{R}}(t,t') & = -\im s_\a(t)s_\a(t') \ex^{-\im\phi_\a(t,t')}\int \frac{\ud\w}{2\pi}\ex^{-\im\w(t-t')} \nonumber \\
& \hspace{50pt}\times 2 \pi \sum_k T_{ik\a}\delta(\w-\eps_{k\a})T_{k\a j} \theta(t-t') \nonumber \\
& = -\frac{\im}{2} s_\a^2(t)\delta(t-t') \Gamma_{\a,ij} ,
\end{align}
where we employed the wide-band limit approximation (WBLA), 
$\Gamma_{\a,ij}(\w)
=2\pi\sum_k T_{i k\a}\delta(\w-\eps_{k\a})T_{k\a j}
\simeq \Gamma_{\a,ij}(\mu)\equiv \Gamma_{\a,ij}$, and we used $\int \frac{\ud\w}{2\pi}\ex^{-\im\w(t-t')}\theta(t-t')=\delta(t-t')/2$. Inserting this into the r.h.s. of Eq.~\eqref{eq:gr} we obtain
\begin{align}
\int \ud \bar{t} \Sigma^{\text{R}}(t,\bar{t}) G^{\text{R}}(\bar{t},t') & = \int \ud \bar{t} \sum_\a [-\im\Gamma_\a s_\a^2(t)/2]\delta(t-\bar{t})G^{\text{R}}(\bar{t},t') \nonumber \\
& = -\frac{\im}{2} \sum_\a \Gamma_\a s_\a^2(t)G^{\text{R}}(t,t'),
\end{align}
so we may write the equation of motion as
\be\label{eq:gret}
\left[\im\frac{\ud}{\ud t} - \heff(t)\right]G^{\text{R}}(t,t') = \delta(t,t')
\ee
with the non-self-adjoint, effective Hamiltonian $\heff(t)\equiv 
h^e(t) - \frac{\im}{2}\sum_\a \Gamma_\a s_\a^2(t)$. When considering 
the GKBA for the lesser and greater Green's functions,  
\be
G^\lessgtr(t,t') = - G^{\text{R}}(t,t')\rho^\lessgtr(t') + \rho^\lessgtr(t)G^{\text{A}}(t,t'),
\ee
we take the quasi-particle propagators for the coupled system as $G^{\text{R}/\text{A}}(t,t') = \mp \im \theta[\pm(t-t')]\mathrm{T} \ex^{-\im\int_{t'}^t \ud \bar{t} \heff(\bar{t})}$. Within the GKBA, the lesser and greater Green's functions thus satisfy~\cite{karlsson_fast_2021supp}
\begin{align}
\im\frac{\ud}{\ud t} G^\lessgtr(t,t') & = \heff (t) G^\lessgtr(t,t'), \\
-\im\frac{\ud}{\ud t'} G^\lessgtr(t,t') & = G^\lessgtr(t,t') \heffd(t') \label{eq:gkba-eom}.
\end{align}

\section{Lesser self-energy}

For the equations of motion for the one-particle density matrix, we also need the lesser self-energy~\cite{tuovinen_time-dependent_2014supp}:
\be
\Sigma_{\a,ij}^<(t,t') = \sum_k T_{ik\a} s_\a(t) g_{k\a}^<(t,t') T_{k\a j}s_\a(t') ,
\ee
where the free-electron, lesser Green's function of the $\a$-th electrode is
\be\label{eq:glss}
g_{k\a}^<(t,t') = \im f(\eps_{k\a}-\mu)\ex^{-\im\int_{t'}^t \ud \bar{t}[\eps_{k\a}+V_\a(\bar{t})]}.
\ee
Here, it is worth noting that the initial condition originates from the Matsubara component, i.e., the Fermi function contains the unbiased electrode energy dispersion shifted by the chemical potential~\cite{svl-booksupp}. Transforming the $k$-summation into a frequency integral gives
\begin{align}
\Sigma_{\a,ij}^<(t,t') & = \im s_\a(t)s_\a(t') \ex^{-\im\phi_\a(t,t')} \int \frac{\ud\w}{2\pi} f(\w-\mu)\ex^{-\im\w(t-t')} \nonumber \\
& \hspace{50pt}\times 2\pi\sum_k T_{ik\a} \delta(\w-\eps_{k\a}) T_{k\a j}.
\end{align}
In matrix form, we may then write (employing again the WBLA)~\cite{Ridley2017supp}
\be\label{eq:sigmalssintegral}
\Sigma_\a^<(t,t') = \im\Gamma_\a s_\a(t)s_\a(t')\ex^{-\im\phi_\a(t,t')}\int\frac{\ud\w}{2\pi}f(\w-\mu)\ex^{-\im\w(t-t')}.
\ee
Away from the time-diagonal, $t\neq t'$, the integral is 
well-behaving, a divergence is arising however for $t=t'$. At zero 
temperature one finds 
$\Sigma^{<}(t,t'\sim t)\sim \frac{1}{t-t'}-i\pi \delta(t-t')$, see 
Eq.~(11) in 
Ref.~\cite{spc.2008supp}.
This divergence is however harmless since in the EOM the 
self-energy is always convoluted with the Green's function over time; see below. 
The time-linear scheme is ``blind'' to the divergence of the 
embedding self-energy since 
it is cast in terms of the embedding correlator, 
which is the integral of the power-law divergent part of 
$\Sigma^{<}$ and the advanced Green's function.

The Fermi function can be evaluated employing a pole expansion~\cite{hu_communication_2010supp, Ridley2017supp}
\be
f(x) \equiv \frac{1}{\ex^{\b x}+1} = \frac{1}{2} - \lim_{N_p\to \infty} \sum_{l=1}^{N_p} \eta_l \left(\frac{1}{\b x+\im\zeta_l}+\frac{1}{\b x -\im\zeta_l}\right) ,
\ee
where $\eta$ and $\pm\im\zeta$ are the residues and poles ($\zeta>0$), respectively. In the Matsubara case, one would take $\eta_l=1$ and $\zeta_l=\pi(2l-1)$, but the expansion can be optimized through the solution of an eigenvalue problem of a specific, tridiagonal matrix~\cite{hu_communication_2010supp}. Because of the exponential $\ex^{-\im\w(t-t')}$, the nontrivial part of the integral is closed on the lower-half complex plane for $t>t'$ and on the upper-half complex plane for $t<t'$:
\begin{align}
& \Sigma_\a^<(t,t') \nonumber \\
& = \im \Gamma_\a s_\a(t)s_\a(t') \ex^{-\im\phi_\a(t,t')} \frac{1}{2}\int\frac{\ud\w}{2\pi}\ex^{-\im\w(t-t')} \nonumber \\
& - \im \Gamma_\a s_\a(t)s_\a(t') \ex^{-\im\phi_\a(t,t')}\sum_l \frac{\eta_l}{\beta} \nonumber \\
& \times \int\frac{\ud\w}{2\pi}\left(\frac{1}{\w-\mu+\im\zeta_l/\beta}+\frac{1}{\w-\mu-\im\zeta_l/\beta}\right)\ex^{-\im\w(t-t')} \nonumber \\
& = \frac{\im}{2}\Gamma_\a s_\a^2(t) \delta(t-t') - \im \Gamma_\a s_\a(t)s_\a(t') \ex^{-\im\phi_\a(t,t')} \sum_l \frac{\eta_l}{\beta}\nonumber \\
& \times \left[-\im\ex^{-\im(\mu-\im\frac{\zeta_l}{\beta})(t-t')}\theta(t-t') +\im\ex^{-\im(\mu+\im\frac{\zeta_l}{\beta})(t-t')}\theta(t'-t)\right], \label{eq:sigmalssintegrated}
\end{align}
where we noticed a missing prefactor $1/\beta$ in Ref.~\cite{Ridley2017supp}. In our situation, we only require the $t>t'$ contribution:
\begin{align}\label{eq:sigmalss}
\Sigma_\a^<(t,t') & = \frac{\im}{2}s_\a^2(t)\delta(t-t')\Gamma_a \nonumber \\
& - s_\a(t)\sum_l \frac{\eta_l}{\beta} s_\a(t') \ex^{-\im\phi(t,t')}\ex^{-\im(\mu-\im\zeta_l/\beta)(t-t')}\Gamma_a .
\end{align}
We observe that the inclusion of finite number of poles removes 
the power-law divergence and makes the function numerically 
more tractable~\cite{Ridley2017supp}.

\section{Embedding integral}
Inserting Eq.~\eqref{eq:sigmaret} into the embedding integral 
appearing in Eq.~(2) of the main text yields
\begin{align}\label{eq:iem}
I_{\text{em}}(t) & = \int \ud \bar{t} \left[\Sigma^<(t,\bar{t})G^{\text{A}}(\bar{t},t)+\Sigma^{\text{R}}(t,\bar{t})G^<(\bar{t},t)\right] \nonumber \\
& = \int \ud \bar{t} \left\{\Sigma^<(t,\bar{t})G^{\text{A}}(\bar{t},t) + \frac{1}{2}\Gamma (t)\delta(t-\bar{t})[-\im G^<(\bar{t},t)]\right\} \nonumber \\
& = \int \ud \bar{t} \Sigma^<(t,\bar{t})G^{\text{A}}(\bar{t},t) + \frac{1}{2}\Gamma(t)\rho(t),
\end{align}
where we defined $\Gamma(t) \equiv \sum_\a s_\a^2(t) \Gamma_\a$ and $\rho(t) \equiv -\im G^<(t,t)$. Further, inserting Eq.~\eqref{eq:sigmalss} gives
\begin{align}\label{eq:emint}
I_{\text{em}}(t) & = \frac{1}{2}\Gamma(t)\rho(t) + \sum_\a \int \ud \bar{t} \left\{ \frac{\im}{2}s_\a^2(t)\delta(t-\bar{t})\Gamma_\a G^{\text{A}}(\bar{t},t) \right.\nonumber\\
& \left.- s_\a(t){\textstyle \sum_l} \frac{\eta_l}{\beta} s_\a(\bar{t}) \ex^{-\im\phi(t,\bar{t})}\ex^{-\im(\mu-\im\zeta_l/\beta)(t-\bar{t})}\Gamma_a G^{\text{A}}(\bar{t},t) \right\} \nonumber \\
& = \frac{1}{2}\Gamma(t)\rho(t) - \frac{1}{4}\Gamma(t) -\sum_{l\a}s_\a(t)\frac{\eta_l}{\beta} \Gamma_\a \mathcal{G}_{l\a}^{\text{em}}(t),
\end{align}
where we introduced the embedding correlator
\be\label{eq:emcorr}
\mathcal{G}_{l\a}^{\text{em}}(t) = \int \ud \bar{t} s_\a(\bar{t})\ex^{-\im\phi_\a(t,\bar{t})}\ex^{-\im\mu(t-\bar{t})}\ex^{-\zeta_l(t-\bar{t})/\beta} G^{\text{A}}(\bar{t},t) .
\ee
In Eq.~\eqref{eq:emint}, the prefactor of the second term results from the implicit step function within the advanced Green's function: $\int \ud \bar{t} \delta(t-\bar{t})G^{\text{A}}(\bar{t},t)=\im/2$.

\section{Equations of motion}

With the above-derived embedding integral, the equation of motion for the electronic density matrix becomes (here we omit the collision integral $I_{\text{c}}$ resulting from, e.g., electron-electron or electron-phonon correlations; it can be directly included)
\begin{align}\label{eq:rhoeom}
& \im \frac{\ud}{\ud t} \rho(t) \nonumber \\
& = \left[ h^e(t) \rho(t) - \im I_{\text{em}}(t)\right] - \text{h.c.} \nonumber \\
& = \left[h^e(t)\rho(t) - \frac{\im}{2}\Gamma(t)\rho(t) + \frac{\im}{4}\Gamma(t) \right.\nonumber\\
& \hspace{20pt}\left. +\im{\textstyle \sum_{l\a}}s_\a(t)\frac{\eta_l}{\beta} \Gamma_\a \mathcal{G}_{l\a}^{\text{em}}(t) \right] - \text{h.c.} \nonumber \\
& = \left[\heff(t)\rho(t) + \frac{\im}{4}\Gamma(t) +\im\sum_{l\a}s_\a(t)\frac{\eta_l}{\beta} \Gamma_\a \mathcal{G}_{l\a}^{\text{em}}(t) \right] - \text{h.c.},
\end{align}
where we utilized the effective Hamiltonian. The equation of motion for the embedding correlator is derived as
\begin{align}\label{eq:emcorreom}
& \im \frac{\ud}{\ud t} \mathcal{G}_{l\a}^{\text{em}}(t) \nonumber \\
& = \int \ud \bar{t} s_\a(\bar{t}) \left\{\im \frac{\ud}{\ud t}\ex^{-\im\int_{\bar{t}}^t \ud \tau \left[V_\a(\tau)+\mu-\im\zeta_l/\beta\right]}\right\}G^{\text{A}}(\bar{t},t) \nonumber \\
& - \int \ud \bar{t} s_\a(\bar{t}) \ex^{-\im\int_{\bar{t}}^t \ud \tau \left[V_\a(\tau)+\mu-\im\zeta_l/\beta\right]} \left\{-\im \frac{\ud}{\ud t}G^{\text{A}}(\bar{t},t)\right\} \nonumber \\
& = \int \ud \bar{t} s_\a(\bar{t}) \left[V_\a(t)+\mu-\im\zeta_l/\beta\right]\ex^{-\im\int_{\bar{t}}^t \ud \tau \left[V_\a(\tau)+\mu-\im\zeta_l/\beta\right]}G^{\text{A}}(\bar{t},t) \nonumber \\
& - \int \ud \bar{t} s_\a(\bar{t}) \ex^{-\im\int_{\bar{t}}^t \ud \tau \left[V_\a(\tau)+\mu-\im\zeta_l/\beta\right]} \left[G^{\text{A}}(\bar{t},t)\heffd(t)+\delta(\bar{t}-t)\right] \nonumber \\
& = -s_\a(t) - \mathcal{G}_{l\a}^{\text{em}}(t)\left[\heffd(t)-V_\a(t)-\mu+\im\zeta_l/\beta\right],
\end{align}
where we used the equation of motion~\eqref{eq:gret} of the quasi-particle propagator for the coupled system $G^{\text{A}}(\bar{t},t)=[G^{\text{R}}(t,\bar{t})]^\dagger$.

The embedding correlator in Eq.~\eqref{eq:emcorr} 
is a $N_{\text{sys}}\times N_{\text{sys}}$ matrix. The 
multiplication in Eq.~\eqref{eq:emcorreom} thus leads to an 
overall computational complexity $
N_{\text{sys}}^3\times N_p \times N_{\rm leads}$.    

\section{$k$-resolved spectral decomposition}

Alternatively, we could also write the embedding integral~\eqref{eq:iem} directly in terms of the $k$-resolved electrode Green's functions (spectral decomposition):
\begin{align}\label{eq:iemk}
I_{\text{em},ij}(t) & = \sum_m\int \ud \bar{t} \left[\Sigma_{im}^<(t,\bar{t})G_{mj}^{\text{A}}(\bar{t},t)+\Sigma_{im}^{\text{R}}(t,\bar{t})G_{mj}^<(\bar{t},t)\right] \nonumber \\
& = \sum_{k\a m} \int \ud \bar{t} \left[ T_{ik\a}s_\a(t) g_{k\a}^<(t,\bar{t})T_{k\a m}s_\a(\bar{t})G_{mj}^{\text{A}}(\bar{t},t) \right. \nonumber \\
& \hspace{50pt} \left. + T_{ik\a}s_\a(t) g_{k\a}^{\text{R}}(t,\bar{t})T_{k\a m}s_\a(\bar{t})G_{mj}^<(\bar{t},t) \right] \nonumber \\
& = \sum_{k\a} T_{ik\a}s_\a(t) \widetilde{\mathcal{G}}_{k\a j}^{\text{em}}(t),
\end{align}
where we introduced another embedding correlator 
\begin{align}\label{eq:emcorrk}
\widetilde{\mathcal{G}}_{k\a j}^{\text{em}}(t) & = \sum_m \int \ud \bar{t} \left[ g_{k\a}^<(t,\bar{t})T_{k\a m}s_\a(\bar{t})G_{mj}^{\text{A}}(\bar{t},t) \right. \nonumber \\
& \hspace{50pt}\left. + g_{k\a}^{\text{R}}(t,\bar{t})T_{k\a m}s_\a(\bar{t})G_{mj}^<(\bar{t},t) \right].
\end{align}
The embedding integral in Eq.~\eqref{eq:iemk} enters the equation of 
motion~(2) in the main text. 
The correlator $\widetilde{\mathcal{G}}_{k\a j}^{\text{em}}$ is a 
scalar quantity and it is distinct from the embedding coorelator 
$\mathcal{G}_{l\a}^{\text{em}}$ in Eq.~\eqref{eq:emcorr}, which 
is a $N_{\text{sys}}\times N_{\text{sys}}$ matrix for every pole $l$ 
and electrode $\a$.

For deriving an equation of motion for $\widetilde{\mathcal{G}}_{k\a 
j}^{\text{em}}(t)$ we need the equations of motion 
for the free-electron, electrode Green's functions, 
cf.~Eqs.~\eqref{eq:gretk} and~\eqref{eq:glss}:      
\begin{align}
\im \frac{\ud}{\ud t} g_{k\a}^{\text{R}}(t,t') & = \bar{\eps}_{k\a}(t)g_{k\a}^{\text{R}}(t,t') + \delta(t-t') , \\
\im \frac{\ud}{\ud t} g_{k\a}^<(t,t') & = \bar{\eps}_{k\a}(t)g_{k\a}^<(t,t'),
\end{align}
where $\bar{\eps}_{k\a}(t) = \eps_{k\a}+V_\a(t)$ is the biased electrode energy dispersion, see below Eq.~\eqref{eq:ham}.
We then find
\begin{align}
& \im \frac{\ud}{\ud t} \widetilde{\mathcal{G}}_{k\a j}^{\text{em}}(t) \nonumber \\
& = \sum_m \int \ud \bar{t} \left\{ \left[ \im \frac{\ud}{\ud t}g_{k\a}^<(t,\bar{t})\right]T_{k\a m}s_\a(\bar{t})G_{mj}^{\text{A}}(\bar{t},t) \right.\nonumber \\
& \hspace{40pt}\left. - g_{k\a}^<(t,\bar{t})T_{k\a m} s_\a(\bar{t})\left[-\im\frac{\ud}{\ud t} G_{mj}^{\text{A}}(\bar{t},t)\right]\right.\nonumber \\
& \hspace{40pt}\left.+ \left[ \im \frac{\ud}{\ud t}g_{k\a}^{\text{R}}(t,\bar{t})\right]T_{k\a m}s_\a(\bar{t})G_{mj}^<(\bar{t},t) \right.\nonumber \\
& \hspace{40pt}\left. - g_{k\a}^{\text{R}}(t,\bar{t})T_{k\a m} s_\a(\bar{t})\left[-\im\frac{\ud}{\ud t} G_{mj}^<(\bar{t},t)\right]\right\} \nonumber \\
& = \sum_m \int \ud \bar{t} \left\{ \bar{\eps}_{k\a}(t)g_{k\a}^<(t,\bar{t}) T_{k\a m}s_\a(\bar{t}) G_{mj}^{\text{A}}(\bar{t},t) \right.\nonumber \\
& \hspace{40pt}\left. -g_{k\a}^<(t,\bar{t}) T_{k\a m}s_\a(\bar{t}) \left[ {\textstyle \sum_p} G_{mp}^{\text{A}}(\bar{t},t)h_{\text{eff},pj}^{e\dagger}(t)+\delta(t-\bar{t}) \right] \right.\nonumber \\
& \hspace{40pt}\left. + \left[ \bar{\eps}_{k\a}(t)g_{k\a}^{\text{R}}(t,\bar{t}) + \delta(t-\bar{t}) \right]T_{k\a m}s_\a(\bar{t})G_{mj}^<(\bar{t},t) \right.\nonumber \\
& \hspace{40pt}\left. - g_{k\a}^{\text{R}}(t,\bar{t})T_{k\a m}s_\a(\bar{t}){\textstyle \sum_p}G_{mp}^<(\bar{t},t)h_{\text{eff},pj}^{e\dagger}(t)\right\} \nonumber \\
& = \bar{\eps}_{k\a}(t)\widetilde{\mathcal{G}}_{k\a j}^{\text{em}}(t) + \im \sum_m T_{k\a m}s_\a(t)\left[\rho_{mj}(t) - f(\eps_{k\a}-\mu)\delta_{mj}\right] \nonumber \\
& - \sum_p\widetilde{\mathcal{G}}_{k\a p}^{\text{em}}(t)h_{\text{eff},pj}^{e\dagger}(t) ,
\label{eq:emcorreomk}
\end{align}
where we employed Eq.~\eqref{eq:glss} for the electrode Green's 
function at the equal-time limit, and the GKBA equation of 
motion~\eqref{eq:gkba-eom} for the lesser Green's function. 
The solution of the EOM for all the scalar quantities  
$\widetilde{\mathcal{G}}_{k\a j}^{\text{em}}(t)$ 
scales like $N_{\text{sys}}^2\times N_k \times N_{\rm leads}$.
The scaling ratio between the pole expansion scheme and the spectral 
decomposition scheme is therefore
 $N_p \times N_{\text{sys}}/N_k$. 

\begin{figure}[t]
\centering
\includegraphics[width=0.475\textwidth]{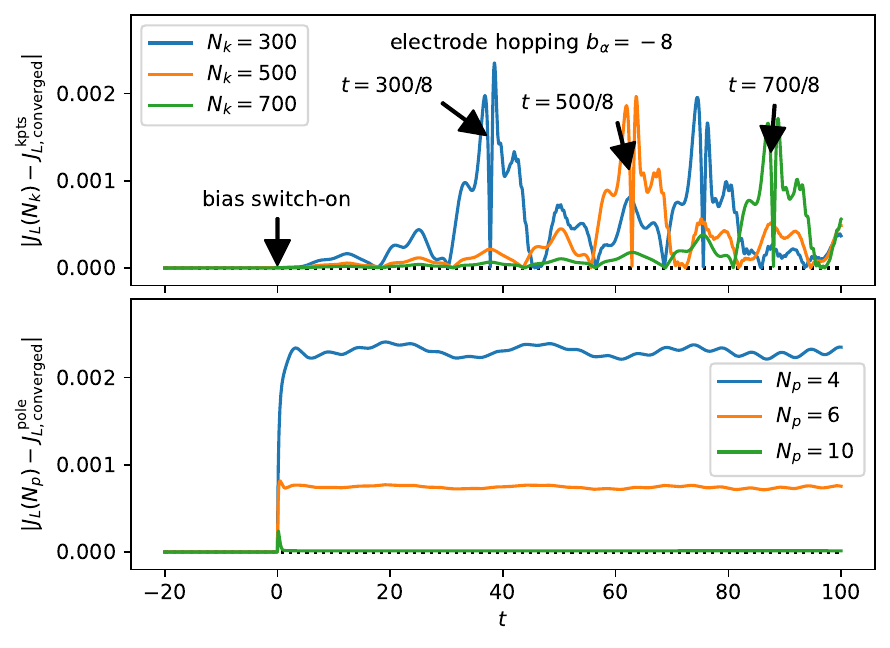}
\caption{Transient currents at the left electrode interface in the dimer case compared to the converged results shown in Fig.~2 of main text. Top: $k$-points spectral decomposition with varying number of points $N_k$. Bottom: pole-expansion with varying number of poles $N_p$.}
\label{fig:comparison}
\end{figure}

In Fig.~\ref{fig:comparison}, we show a calculation corresponding to 
the dimer system studied in Fig.~2 of the main text. The system is coupled to one-dimensional tight-binding electrodes
(on-site energy $a_\a=\mu$, hopping $b_\a$) with the energy 
dispersion $\e_{k\a}=a_\a - 2|b_\a|\cos[\pi 
k/(N_k+1)]$ and the tunneling matrix elements (to terminal sites of the electrodes) $T_{ik\a}=T_\a\sqrt{2/(N_k+1)}\sin(\pi k/(N_k+1))$, where $N_{k}$ is the number of discretized $k$ points. Here, the 
left-interface current, evaluated for different numbers of the $k$ 
points ($N_k$) or poles ($N_p$) is compared to the converged result 
shown in the main text. We see that a recurrence time due to a 
finite-size effect is present in the spectral 
decomposition scheme. This 
recurrence time is equal to $N_k/|b_\alpha|$ and it corresponds to the 
time it takes for the electronic wavefront to go from one of the 
interfaces to the boundary of the corresponding electrode and back.   
 Other limitations of 
the spectral decomposition scheme are discussed in the 
main text. In contrast, the pole expansion scheme shows no finite-size 
effects, but instead, if the number of poles is too low, the 
steady-value of the current is inaccurate. 
Compared to the  
spectral decomposition scheme, the pole expansion scheme 
converges extremely fast.

\begin{figure}[b]
\centering
\includegraphics[width=0.425\textwidth]{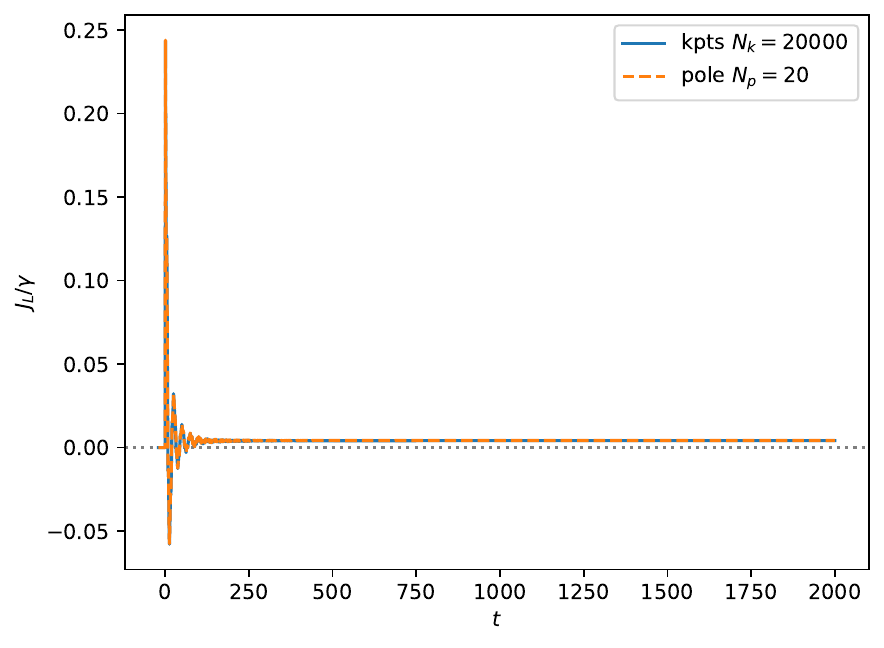}
\caption{Long-time behaviour corresponding to Fig. 2(b) of the main text using the
spectral-decomposition scheme and the pole expansion of the Fermi function.}
\label{figS1}
\end{figure}

Within the temporal window of Fig.~2 in the main text and Fig.~\ref{fig:comparison} above (up to $t=100$), the 
current has not yet reached a steady value. By evolving longer in time, 
however, a well defined steady state is attained. 
In Fig.~\ref{figS1} we show the results of a longer time evolution (up to 
$t=2000$); the current saturation is clearly visible.   
For this long-time evolution, a significantly higher number of $k$-points 
is required to reach converged results, in contrast to the number of 
poles which is instead the same.

\section{Meir-Wingreen formula}

The current between the system and the $\a$-th electrode can be calculated from the Meir-Wingreen formula, which consists of the $\a$-th electrode contribution to the embedding integral, cf.~Eq.~\eqref{eq:emint}:
\begin{align}
& J_\a(t) = 2 \mathrm{Re}\mathrm{Tr} I_\a(t) \nonumber \\
& = 2 \mathrm{Re}\mathrm{Tr} \left[ \frac{1}{2}\Gamma_\a s_\a(t)\rho(t) - \frac{1}{4}\Gamma_\a s_\a(t) -\sum_l s_\a(t)\frac{\eta_l}{\beta} \Gamma_\a \mathcal{G}_{l\a}^{\text{em}}(t) \right] \nonumber \\
& = 2s_\a (t)\mathrm{Re}\mathrm{Tr}\left[\Gamma_\a \left( \frac{2\rho(t)-1}{4} -{ \sum_l}\frac{\eta_l}{\beta}G_{l\a}^{\text{em}}(t) \right)\right] ,
\end{align}
where the trace also contains a sum over spin.

Alternatively, the Meir-Wingreen formula can be cast in terms of the $k$-resolved embedding correlator:
\be
J_\a(t) = 2\mathrm{Re} \sum_{ik} T_{ik\a}s_\a(t) \widetilde{\mathcal{G}}_{k\a i}^{\text{em}}(t) .
\ee

Remarkably, calculating the time-dependent current adds no extra 
complexity in either cases. The current formula is completely 
specified in terms of the single-time embedding correlator, which is 
readily available when evolving the coupled system of ODEs. With the 
pole expansion, this corresponds to Eqs.~\eqref{eq:rhoeom} 
and~\eqref{eq:emcorreom}, and with the $k$-resolved spectral 
decomposition, to the first equality of Eq.~\eqref{eq:rhoeom}, and 
Eqs.~\eqref{eq:iemk} and~\eqref{eq:emcorreomk}.

\end{document}